\documentstyle[prb,aps,epsf,psfig]{revtex}
\begin{document}
 \draft

\title{Realistic Tight Binding Model for the Electronic Structure of II-VI Semiconductors}
\author{Sameer Sapra, N. Shanthi and D.D. Sarma$^*$}
\address{\it Solid State and Structural Chemistry Unit,
Indian Institute of Science, Bangalore - 560012, India}
\date{\today}

\maketitle

\begin{abstract}
We analyze the electronic structure of group II-VI semiconductors
obtained within LMTO approach in order to arrive at a realistic
and minimal tight binding model, parameterized to provide an
accurate description of both valence and conduction bands. It is
shown that a nearest-neighbor $sp^3d^5$ model is fairly sufficient
to describe to a large extent the electronic structure of these
systems over a wide energy range, obviating the use of any
fictitious $s^*$ orbital. The obtained hopping parameters obey the
universal scaling law proposed by Harrison, ensuring
transferability to other systems. Furthermore, we show that
certain subtle features in the bonding of these compounds require
the inclusion of anion-anion interactions in addition to the
nearest-neighbor cation-anion interactions.
\end{abstract}
\pacs{PACS Nos. 71.20.-b, 71.20.Nr}
\newpage

\section{Introduction}
The tight binding (TB) method has been employed extensively during
the last few decades for the study of tetrahedrally coordinated
semiconductors, due to the simplicity of the approach and its
ability to describe properties in terms of chemical bonds; this
gives the model a more realistic nature as opposed to methods
based on weak periodic potentials.~\cite{SK,chadi,harrison} The TB
approach is suitable to handle larger systems compared to methods
based on plane waves, due to the low computational costs. The TB
method was originally described by Slater and Koster as an
interpolation scheme.~\cite{SK} It has been developed extensively
since then and is now a well established technique to elucidate
the electronic structure of solids.~\cite{harrison}

For tetrahedral semiconductors, chemical intuition leads one to
consider a minimal $sp^3$ basis on various kinds of atoms in the
solid, and interactions only between the nearest-neighbor atoms
seem necessary. Such a model describes the valence band electronic
structure with a limited accuracy, however, it is now
well-established that such a minimal model cannot reproduce the
band gap,~\cite{chadi,harrison} performing even worse in
describing the overall conduction band electronic structure. In
order to obtain an accurate estimate of the bandgap, Vogl and
coworkers~\cite{vogl} added a fictitious $s^*$ orbital to the
$sp^3$ basis. Adjusting the various electronic parameters for the
$sp^3s^*$ TB model, it was possible to simulate the conduction and
valence band extremal energies, thereby yielding the correct
bandgap. However, this approach failed to account for the band
dispersions even for the lowest unoccupied band. This failure is
not surprising in view of the fact that the inclusion of the $s^*$
orbital and the associated electronic parameter strengths are
merely {\em ad hoc} parameters without any rigorous physical
basis. However, almost all efforts in obtaining TB
parameterization to describe the electronic structures of such
semiconductors have proceeded along these lines; calculations for
a number of tetrahedral semiconductors have resulted in the
establishment of a {\em universal} $sp^3s^*$ model based on
Harrison's $d^{-2}$ law for the interatomic matrix elements of the
TB Hamiltonian.~\cite{vogl} The {\em universal} model is useful as
only the interatomic distances are required to obtain the
interaction parameters, but its applicability is limited as it
does not give a good decription of the unocupied part, {\it e.g.}
as discussed in the case of GaP.~\cite{vogl} Further improvements
in the TB model were achieved by incorporating $d$ orbitals in the
basis.~\cite{richardson,chang} Recently, a TB model based
on the $sp^3d^5s^*$ basis set was employed for the group IV and
III-V semiconductors.~\cite{bassani} This empirical model based on
the nearest-neighbor interactions gives a good description of the
electronic structure of these semiconductors, especially at
high-symmetry points.

The need for a physical and accurate parametrization capable of
describing both valence and conduction bands, and not merely the
bandgap, of these semiconductors is evident. There are direct
experimental probes such as the photoemission and inverse
photoemission that map out the density of states (DOS) of the
valence and conduction band regions. An analysis of such
experiments requires a suitable TB parameterization that work
equally well for the occupied as well as the unoccupied states.
Furthermore, it is possible to obtain experimental information on
the partial density of states by using the site and angular
momentum specific x-ray emission and absorption experiments; it is
then desirable to have a TB model excluding the fictitious $s^*$
orbital. Additionally, we have recently found~\cite{ranjani} in
the context of InP that even $sp^3d^5s^*$ TB parameterization does
not work very well to describe the changes in the electronic
structure as a consequence of spatial localization in nanometric
clusters. Thus, it appears highly desirable to analyze the
electronic structures of such semiconductors and thereby construct
a physical as well as minimal model that would work satisfactorily
for all these diverse cases.

In order to achieve this, we first study the Linearized Muffin Tin
Orbital (LMTO) method~\cite{LMTO1,LMTO2} derived density of states
(DOS), partial density of
states (PDOS) and the crystal orbital Hamiltonian overlap (COHP)
to establish the relative importance of various orbitals in
bonding as well as in determining the details of the electronic
structure in different energy regions. This helps us to identify
the important orbitals. From this
analysis we construct the minimal model, without the $s^*$
orbital. Since we construct the final TB model in successive steps
of including various interactions, we understand in detail the
influence of each of these improvements to modifying the energy
dispersions of various bands. Our final results for the II-VI
semiconductors using the $sp^3d^5$ orbital basis are in excellent
agreement with the LMTO calculations and the various interaction
parameters obtained here obey the universal scaling law.

\section{Methodology}

The band structures of the A$^{\rm II}$B$^{\rm VI}$ type
semiconductors, where A~=~Zn, Cd, Hg and B~=~S, Se, Te, are
calculated using the Linearized Muffin Tin Orbital (LMTO) method
in the Atomic Sphere Approximation (ASA). The zinc-blende
structure, which has one formula unit of A$^{\rm II}$B$^{\rm VI}$
per unit cell, has been studied. The basis set of $s$, $p$ and
$d$ orbitals was used for both the cation and the anion for all
the compounds. Empty spheres were introduced in all cases in order
to keep the overlap of atomic spheres within 16\% in every case.
Only $s$ orbital is used for the empty spheres. The
self-consistency was achieved with 28~$k$-points in the
irreducible Brillouin Zone and band dispersions and density of
states were obtained in each case.

In order to obtain a detailed understanding of the origin of
various features in the electronic structure, we also calculate
the partial densities of states corresponding to cation and anion
$s$, $p$ and $d$ states. While the partial densities of states
provide us with the information concerning the relative
contributions of various orbitals at different energy regions, it
cannot provide any $k$-dependent information. In order to obtain
such momentum-related information, we have additionally analyzed
the orbital character of the band eigen states and shall present
these in terms of the so called ``fatband'' representation of the
band dispersions. However, such analysis does not provide an
insight on the range of interactions important for the system. The
range of interaction is one of the most important ingredients to
determine the suitable tight binding model, as it dictates whether
a nearest-neighbor-only model is sufficient or there is a need to
include farther neighbor interactions. This issue can be addressed
by computing the crystal orbital Hamiltonian population (COHP) for
various pairs of orbitals and atoms, as it provides the relative
contributions to bonding arising from different interactions in
the system.~\cite{COHP}

The tight binding calculations were performed using the
Hamiltonian

\begin{eqnarray}
{\bf H}&=& \sum_{i l_1 \sigma} \epsilon_{l_1} a^{\dagger}_{i l_1 \sigma}a_{i l_1 \sigma}+\sum_{ij}
 \sum_{l_1, l_2,\sigma}(t^{l_1 l_2}_{ij}a^{\dagger}_{i {l_1} \sigma}a_{j {l_2} \sigma} + {\it h.c.})
\end{eqnarray}
where, the electron with spin ${\sigma}$
is able to hop from the
orbitals labelled $l_1$ with onsite energies equal
to $\epsilon_{l_1}$
in the $i^{th}$ unit cell to those labelled
$l_2$ in the $j^{th}$ unit cell, with the summations over
$l_1$ and $l_2$ running over all the orbitals considered
on the atoms in a unit cell,
and $i$ and $j$ over all the unit cells in the solid. Thus,
any orbital in the solid can be defined with the two indices, $i$ and $l_1$.
The  hopping interaction strength ($t^{l_1 l_2}_{ij}$) depends on
the nature of the orbitals involved as well as on the geometry of
the lattice.~\cite{SK} To start with, we estimate the values of
the various onsite energies ($\epsilon$'s) and hopping
interactions ($t$'s) from the LMTO band dispersions and the
density of states. Then, a least-squared-error fitting is carried
out by varying the $\epsilon$'s and $t$'s, calculating the band
dispersions at a number of high symmetry points and then comparing
with the LMTO band dispersions. In the following section, we
present the detailed analysis with the help of ZnS as an
illustrative example; the results obtained from all other systems
are very similar.

\section{Results and Discussion}

In Fig.~\ref{sp3}a, we show the LMTO derived band dispersions for
ZnS along various symmetry directions. The lowest lying band at
about -12~eV is due to S $s$ states, while the group of five flat
bands near -6.4~eV arises from Zn $d$ states. The main part of the
valence band region in ZnS appearing between -5.4 and 0~eV is
contributed by three strongly dispersing bands arising primarily
from the S $p$ states. The lowest lying conduction band centered
around 4~eV is nominally the Zn $s$ derived band, while the next
three bands are attributed to Zn $p$ states. The parts of bands
appearing at the top of the figure are contributed dominantly by
higher lying states, such as the S $d$ levels. The band gap
appearing at the $\Gamma$ point is about 3.2~eV in this
calculation. These results are consistent with previously
published band structure of ZnS.~\cite{znsbands}
a
These results suggest that main parts of the valence and the
conduction bands in ZnS across the  band gap are essentially due
to Zn $s$, $p$ and S $s$, $p$ states, suggesting a TB model
consisting only of these levels as the simplest possible starting
point.This point of view also makes chemical sense as the
tetrahedral coordination around both Zn and S can be easily
achieved in terms of the $sp^3$ hybrid orbitals. However, we show
that such a simplistic model performs very poorly in describing
the electronic structure. For this purpose, we carried out a
detailed fitting of the six corresponding LMTO bands in terms of a
nearest-neighbor TB model with the $sp^3$ basis. The resulting
best description is shown in Fig.~\ref{sp3}b in terms of the TB
band dispersions with the optimized TB parameters. The Zn $d$
bands near -6.4~eV and high lying S $d$ bands are naturally
missing in the TB results. We find that the low lying S $s$ band
is reasonably well described in this simplest model. However, the
TB band dispersions for both the valence and the conduction bands
are considerably different from those in the LMTO calculation. For
example, the band dispersions along X~-~W~-~L~-~K~-~$\Gamma$
within the valence band region are drastically different between
the two calculations. Moreover, not only the band gap is
substantially wrong in the TB results, the curvature of the lowest
lying conduction band near the $\Gamma$ point is very poorly
described within the TB model. The results clearly suggest the
need to go beyond the simplest $sp^3$ nearest-neighbor TB model to
provide a realistic description of the electronic structure of ZnS.

In order to understand the origin of these discrepancies, we plot
the total as well as various partial DOS of ZnS in
Fig.~\ref{znspdos}, with the Zn related partial DOS in
Fig.~\ref{znspdos}a and those related to the S site in
Fig.~\ref{znspdos}b. Focussing on the energy region for the
discrepancies discussed above, we note that the valence band
features appearing between -5.4 and 0~eV are indeed dominated by S
$p$ states (Fig.~\ref{znspdos}b); however, these states have
substantial admixture from the Zn $p$ and $d$ states
(Fig.~\ref{znspdos}a). Since the band formation in a
nearest-neighbor model is entirely due to S - Zn interactions, it
is obvious that Zn $d$ states, contributing as much as the Zn $p$
states in the formation of the valence band, cannot be left out of
any realistic description of the valence band region of ZnS.
Likewise, it is evident in the results for the conduction band
region in Fig.~\ref{znspdos}, particularly in the energy region
approximately between 7 and 12~eV, that the S $d$ contributions
are almost dominant. This must arise from very large Zn $p$ - S
$d$ interactions in forming the upper part of the conduction band
region, establishing the need to include the S $d$ states also in
the TB basis for a satisfactory description of the electronic
structure comprising both valence and conduction band regions.

In order to obtain a more detailed understanding, as well as
insight in the momentum-specific discrepancies, we present the
LMTO band dispersions along the symmetry lines in the fatband
representation in six different panels in Fig.~\ref{znsfb}. While
the band dispersions in each of these six panels are identical,
the width (or the ``fatness'') associated with each band at every
$k$-point is proportional to the orbital character represented in
that panel; for example, Fig.~\ref{znsfb}a shows the contribution
of Zn $s$ states to each of the band eigen-states. These results
clearly establish the detailed nature of the band states. For
example, the band dispersion near -12~eV is dominated by S $s$
states (Fig.~\ref{znsfb}d), formed via the interactions with Zn
$s$, $p$ and $d$ states (see Fig.~\ref{znsfb}a-c). Likewise, the
flat bands near -6.4~eV are primarily Zn $d$ bands
(Fig.~\ref{znsfb}c) formed via the interactions with the S $p$
states (Fig.~\ref{znsfb}e). More importantly, the three strongly
dispersing bands in the valence band region have the S $p$
character (Fig.~\ref{znsfb}e), formed due to substantial S $p$ -
Zn $s$, $p$ and $d$ interactions (Figs.~\ref{znsfb}a-c),
confirming the essential role played by Zn $d$ states in
determining the valence band electronic structure. Likewise, the
extensive S $d$ contributions in all the conduction band states
are also evident in Fig.~\ref{znsfb}f. The inability of the $sp^3$
model to describe the curvature of the lowest conduction band near the $\Gamma$
point (Fig.~\ref{sp3}) can also be understood in terms of these
fatbands. The band state at the $\Gamma$ point is composed of Zn
$s$ admixed with S $s$ states; however, these band states acquire
rapidly changing contributions from S $p$ and $d$ states as $k$
moves away from the $\Gamma$ point, affecting the detail of the
band dispersion in this region of the momentum space.

The above analysis clearly points to the need of including both Zn
and S $d$ states in the basis of the TB model for a realistic
description of the valence and conduction band electronic
structures of ZnS. While we have presented here the detailed
analysis for only the case of ZnS, we carried out similar analysis
for all the compounds and arriving at the same conclusion
concerning the importance of cationic and anionic $d$ states.
Therefore, we carried out a detailed least-squared-error fitting
of the LMTO derived band dispersions in terms of the TB
dispersions with $sp^3d^5$ basis as a function of all the
electronic parameters (on-site and hopping energies) appearing in
the TB Hamiltonian. The fitting was carried out in two successive
steps. First, we performed a fitting of all the eighteen bands
arising primarily from Zn and S $s$, $p$ and $d$ states, though
the S $d$ derived bands appearing at a very high energy above the
Fermi energy do not have any significant bearing on electronic,
optical or chemical properties of the system. However, the
inclusion of the S $d$ derived band dispersions in the first step
of fitting ensures that we use a realistic and physically sound
value for the S $d$ site energy. We then fix the S $d$ site energy
to this value in the second step of the fitting and re-optimize
the other electronic parameters to arrive at the best description
for the thirteen lowest bands with primarily Zn $s$, $p$, $d$ and
S $s$, $p$ characters. The results of this optimization process
are tabulated in Table~I for all the compounds studied here, while
the best-fit TB dispersions within this $sp^3d^5$ nearest-neighbor
model for ZnS are compared with the $ab-initio$ LMTO dispersions
in Fig.~\ref{dispnnonly}. The improvement in using the $sp^3d^5$
model compared to the results obtained from $sp^3$ model
(Fig.~\ref{sp3}) is evident in Fig.~\ref{dispnnonly}. We find that
all the band dispersions, covering both the valence and conduction
bands, as well as the curvature of the lowest conduction band near
the $\Gamma$ point are almost satisfactorily described. We believe
that these parametrizations, , summarized in Table~I, should
already be useful in modelling these semiconductors to a large
extent. However, we can still notice certain discrepancies between
the LMTO and TB dispersions given in Fig.~\ref{dispnnonly}. The
major deviations of the TB band dispersions from the LMTO ones are
marked by rectangular boxes in Fig.~\ref{dispnnonly}.

In order to understand the origin of these discrepancies, we show
crystal orbital Hamiltonian population (COHP) analysis for ZnS in
Fig.~\ref{znscohp}. The total COHP is compared with contributions
arising from Zn-S interactions in Fig.~\ref{znscohp}a. This
clearly shows that the total COHP deviates significantly from that
arising from Zn-S nearest-neighbor interactions alone, suggesting
a longer range interaction also playing a significant role in
bonding. We show the COHP contributions arising from S-S and Zn-Zn
interactions in Figs.~\ref{znscohp}b and c respectively. These
results clearly show that while Zn-Zn interaction
(Fig.~\ref{znscohp}c) is small and can possibly be neglected, S-S
interaction (Fig.~\ref{znscohp}b) contributes significantly and is
often comparable to Zn-S interactions in certain energy ranges.
Thus, it is evident that a more accurate description of the
electronic structure of ZnS must include next-nearest-neighbor S-S
interactions along with the nearest-neighbor Zn-S interactions.
Thus, we carried out a detailed fitting of the LMTO band
dispersions in terms of a TB model in the $sp^3d^5$ basis, as
before, but including the next-nearest-neighbor S-S interactions
along with the nearest-neighbor Zn-S interactions. We follow the
same two step approach to the fitting, as described before. The
resulting TB parameters for the best fit results for each compound
are given in Table~II and an illustrative example of the simulated
band dispersions are shown in Fig.~\ref{znscdisp} using the case
of ZnS. Most of the deviations in the band dispersions observed in
the case of the nearest-neighbor model (Fig.~\ref{dispnnonly}) are
largely removed, in the results shown in Fig.~\ref{znscdisp},
leading to an excellent agreement with the $ab-initio$ results. We
have further verified the reliability of these parameters in
describing, not only the band dispersions along the symmetry
directions, but also the overall electronic structures by
computing the density of states within the TB model. In
Fig.~\ref{znscdos}, we show the comparison of DOS obtained from
LMTO and that from the present TB model for the case of ZnS over
the valence and conduction band regions. The figure shows a very
good agreement between the two.

An important step in demonstrating the usefulness of such
parameterized tight binding approaches was realized by the Bond
Orbital Model proposed by Harrison~\cite{harrison} who showed that
the hopping interaction strengths follow a universal scaling law
with the distance between the two orbital sites. For a large
number of systems, it was shown that the Slater-Koster parameters
have a dependence of $d^{-2}$, where $d$ is the distance between
the two sites connected by the hopping integral. This observation
ensures that the extracted parameters are transferable to other
crystal structures. This approach was further extended~\cite{vogl}
to include a description of the lowest conduction band along with
the valence band within a nearest-neighbor $sp^3s^*$ model.

In order to establish the transferability, and consequently the
usefulness, of the hopping parameter values obtained here, we have
examined the scaling behavior of these parameters. In
Fig.~\ref{nnscale}, we plot the various hopping interaction
strengths (SKK) obtained within the nearest-neighbor-only model
(Table~I) multiplied by $d^2$ as a function of $d$ for all the
compounds. This figure clearly shows that the parameter values
follow the $d^{-2}$ scaling law reasonably well, with SKK*$d^2$
being nearly independent of $d$ for each type of hopping parameters,
as shown in the figure by the
horizontal lines representing the average SKK*$d^2$ values which are
also listed in Table~I. We find that the primary deviations from
the scaling laws are for the
three compounds with Te, namely ZnTe, HgTe and
CdTe, while the compounds of S and Se obey the scaling law considerably.
It turns out that even for
the tellurides, the deviations from the scaling law do not
significantly vitiate the description of the electronic structure.
To illustrate this point, we show in Fig.~\ref{nncomp} the band
dispersions obtained for ZnTe, which exhibits one of the largest
deviations from the expected scaling law. Fig.~\ref{nncomp}a
describes the $ab-initio$ band dispersions obtained within LMTO,
while Fig.~\ref{nncomp}b shows the best fit obtained with tight
binding nearest-neighbor-only model. The corresponding parameter
values are given in Table~I and the data points are plotted in
Fig.~\ref{nnscale} multiplied by the corresponding $d^2$. The band
dispersions obtained within the same model, but with parameter
values corresponding to the average (SKK*$d^2$) instead of the
best-fit optimized parameters, are shown in Fig.~\ref{nncomp}c. A
comparison of Fig.~\ref{nncomp}b and c shows hardly any difference
between the two, both providing excellent description of the
$ab-initio$ band dispersions shown in Fig.~\ref{nncomp}a. This
indicates that the parameters obtained within the
nearest-neighbor-only model are transferable to other cases.

We have also examined the scaling behavior of the hopping parameters obtained with the next
nearest-neighbor model (Table~II) and found a similar $d^{-2}$ dependence. The corresponding
average values of SKK*$d^2$ are also given in Table~II. Using these average
values of the hopping parameters we have calculated the band dispersions for ZnTe, shown in
Fig.~\ref{nncomp}d. These band dispersions also provide an excellent description of the band
dispersions obtained within the $ab-initio$ approach, shown in Fig.~\ref{nncomp}a.

\section{Conclusion}

In conclusion, we have presented a systematic development of parameterized tight binding
model for the description of the electronic structures of group II-VI semiconductors
comprising both the valence and the conduction bands. We analyze the nature and origin
of bonding as well as the atomic orbital contributions to each band
eigen-states to arrive at the necessary minimal model involving $sp^3d^5$ orbitals at the
cationic and the anionic sites, obviating the need to use any fictitious $s^*$ orbital in the
basis. Even a nearest-neighbor-only model is found to provide an excellent description of
the $ab-initio$ band dispersions and the density of states over a wide energy range
covering the entire valence and conduction band regions. The obtained hopping parameters
are shown to observe the $d^{-2}$ scaling law of the Bond Orbital Model proposed by Harrison.
Furthermore, we also obtain the parameter
values in a next nearest-neighbor tight binding model that further improves the agreement of
this empirical approach to the $ab-initio$ results, capturing some subtle features in bonding
in these compounds, particularly involving the top of the valence band.

\newpage
\begin{centering}
FIGURE CAPTIONS
\end{centering}

\begin{figure}
\caption{\label{sp3} Band dispersions for the zinc blende ZnS (a) LMTO results using the
$s$, $p$ and $d$ orbital basis on both Zn and S, (b) the tight binding results for $sp^3$
orbital basis fit to the LMTO results. The zero of the energy scale is set at the top of
the valence band.}
\end{figure}
\begin{figure}
\caption{\label{znspdos}Density of states and partial density of states for zinc-blende ZnS
calculated using LMTO-ASA method: (a) Zn $s$, $p$ and $d$ PDOS; (b) S $s$, $p$ and $d$ PDOS}
\end{figure}
\begin{figure}
\caption{\label{znsfb}Band dispersions for the zinc-blende ZnS showing fatbands for
(a)Zn $s$, (b) Zn $p$, (c) Zn $d$, (d)S $s$, (e) S $p$ and (f) S $d$.}
\end{figure}
\begin{figure}
\caption{\label{dispnnonly} Band dispersions for zinc blende ZnS: (a) LMTO results and
(b) tight binding fitted results for the nearest-neighbor interactions only in the
$sp^3d^5$ orbital basis on both Zn and S.}
\end{figure}
\begin{figure}
\caption{\label{znscohp}COHP for zinc-blende ZnS. Top panel shows the total COHP alongwith
the Zn-S interaction COHP. Middle panel shows the COHP for S-S interaction
and the bottom panel contains the COHP for Zn-Zn interaction.}
\end{figure}
\begin{figure}
\caption{\label{znscdisp}Comparison of band dispersions for ZnS from (a) LMTO
and (b) tight-binding fit with the $sp^3d^5$ orbital basis on both Zn and S. Both
Zn-S and S-S interactions are included in the TB model.}
\end{figure}
\begin{figure}
\caption{\label{znscdos}Comparison of total density of states for
ZnS from (a) LMTO and (b)tight-binding fit with the $sp^3d^5$ orbital basis on
both Zn and S. Both Zn-S and S-S interactions are included in the
TB model.}
\end{figure}
\begin{figure}
\caption{\label{nnscale}SKK*$d^2$ versus $d$ for all the II-VI semiconductors studied
using the TB model with the $sp^3d^5$ basis on both the anion and the cation with only
nearest-neighbor interactions; SKK represents the various hopping parameters and
$d$ is the distance between the cation and the anion. The plot
establishes the scaling behavior of the hopping parameters $ss\sigma$, $sp\sigma$, $pp\sigma$,
$pp\pi$ and $ps\sigma$.}
\end{figure}
\begin{figure}
\caption{\label{nncomp}Comparison of band dispersions for ZnTe obtained from
(a) LMTO, (b) TB-fit using the $sp^3d^5$ orbital basis with nearest-neighbor-only
interactions, (c) scaling parameters obtained from the above model. The parameters
are extracted from the SKK*$d^2$ values in Table~I. (d)
scaling parameters obtained from the TB-fit model with the
anion-anion next nearest-neighbor interactions (values from Table~II).}
\end{figure}

\newpage
~~~~~~~~~~~~~~~~~~~~~~~~~~~~~~~~~~~~~~~~~~~~Table~I\\
Tight-binding parameters obtained from the least-square-error fit to LMTO
band dispersions for the nine II-VI semiconductors in the $sp^3d^5$ basis with only the
nearest-neighbor Zn-S interactions.The first row lists the interatomic spacings in {\AA},
the next eight rows contain the onsite energies for all the orbitals, {\it e.g.} the row
for $d_c(t_2)$ lists the entries for the the $t_2$ $d$ orbital onsite energies
for the cation. The subscript $a$ denotes the anion. The last eleven rows list
the Slater Koster parameters.~\cite{note}
The last column shows
the average value of the Slater Koster parameters multiplied by the square
of the cation-anion distance, $d^2$.
\vspace*{1cm}
\begin{table}[b]
\begin{centering}
{\small
\begin{tabular}{|l|cccccccccc|}
\hline
&&&&&&&&&&\\
&~~ZnS~~&~~ZnSe~~&~~ZnTe~~&~~CdS~~&~~CdSe~~&~~CdTe~~&~~HgS~~&~~HgSe~~&~~HgTe~~&Average\\
&&&&&&&&&&$SKK*d^2$\\
\hline
&&&&&&&&&&\\
$d$ (\AA)&2.34&2.45&2.64&2.52&2.62&2.81&2.53& 2.63&2.80&\\
&&&&&&&&&&\\
$s_c$&0.92&0.74&1.06&0.47&0.30&0.40&-2.03&-2.05&-1.75&\\
$p_c$&8.40&8.38&7.24&7.94&7.91&7.21&7.89&7.68&7.12&\\
$d_c(t_2)$&-5.82&-6.16&-6.92&-6.83&-7.31&-7.97&-5.99&-6.15&-6.91088&\\
$d_c(e)$&-6.21&-6.47&-7.24&-7.44&-7.81&-8.42&-6.31&-6.53&-7.21&\\
$s_a$&-10.33&-11.24&-10.68&-10.58&-11.45&-10.38&-11.04&-11.57&-10.49&\\
$p_a$&2.41&1.93&2.29&1.43&0.93&1.13&0.69&0.66&0.03&\\
$d_a(t_2)$&15.54&16.48&13.23&14.43&15.26&12.83&14.84&15.68&12.8292&\\
$d_a(e)$&13.60&14.48&12.23&13.15&14.10&11.63&12.87&13.54&11.66&\\
&&&&&&&&&&\\
$ss\sigma$&-1.35&-1.01&-0.54&-1.02&-0.74&-0.47&-1.07&-0.93&-0.74&-5.73\\
$sp\sigma$&2.45&2.33&2.26&2.12&2.06&1.93&1.92&1.85&1.82&13.95\\
$pp\sigma$&4.76&4.37&4.01&4.18&3.70&3.54&3.93&3.75&3.08&26.18\\
$pp\pi$&-0.84&-0.83&-0.95&-0.64&-0.67&-0.69&-0.69&-0.71&-0.86&-5.17\\
$ps\sigma$&-2.25&-1.89&-0.52&-1.99&-1.66&-0.94&-1.81&-1.87&-1.25&-10.36\\
$ds\sigma$&-0.05&-0.04&-0.00&-0.00&-0.01&-0.26&-0.72&-0.54&-0.52&-1.67\\
$dp\sigma$&1.37&1.19&1.29&1.75&1.52&1.52&1.45&1.42&1.29&9.61\\
$dp\pi$&-0.45&-0.39&-0.34&-0.35&-0.32&-0.27&-0.59&-0.50&-0.31&-2.61\\
$sd\sigma$&-2.59&-2.71&-3.05&-1.14&-1.22&-2.11&-0.94&-1.40&-1.79&-12.65\\
$pd\sigma$&-2.78&-2.78&-3.36&-1.29&-1.09&-2.42&-1.37&-1.45&-1.73&-13.62\\
$pd\pi$&2.31&2.42&2.27&2.15&2.38&1.90&2.014&2.21&1.82&14.49\\
\hline
\end{tabular}
}
\end{centering}
\end{table}
\newpage
~~~~~~~~~~~~~~~~~~~~~~~~~~~~~~~~~~~~~~~~~~~~Table~II\\
\begin{table}[b]
Tight-binding parameters obtained from the least-square-error fit to LMTO
band dispersions for the nine II-VI semiconductors in the $sp^3d^5$ basis with the
Zn-S and S-S interactions. The first row lists the interatomic spacings in {\AA},
the next eight rows contain the onsite energies for all the orbitals, {\it e.g.} the row
for $d_c(t_2)$ lists the entries for the the $t_2$ $d$ orbital onsite energies
for the cation. The subscript $a$ denotes the anion. The last fifteen rows list
the Slater Koster parameters.~\cite{note}
The last column shows
the average value of the Slater Koster parameters multiplied by the square
of the cation-anion distance, $d^2$.
\vspace*{2cm}
\begin{centering}
{\small
\begin{tabular}{|l|cccccccccc|}
\hline
&&&&&&&&&&\\
&~~ZnS~~&~~ZnSe~~&~~ZnTe~~&~~CdS~~&~~CdSe~~&~~CdTe~~&~~HgS~~&~~HgSe~~&~~HgTe~~&Average~\\
&&&&&&&&&&$SKK*d^2$\\
\hline
&&&&&&&&&&\\
$d$ (\AA)&2.34&2.45&2.64&2.52&2.62&2.81&2.53&2.63&2.80&\\
&&&&&&&&&&\\
$s_c$&2.29&1.40&0.50&1.44&0.83&0.50&-0.85&-0.95&-1.45&\\
$p_c$&9.32&9.21&8.36&8.22&7.80&7.78&8.16&8.35&7.76&\\
$d_c(t_2)$&-6.24&-6.46&-7.26&-7.53&-7.85&-8.46&-6.10&-6.66&-7.27&\\
$d_c(e)$&-6.16&-6.40&-7.21&-7.38&-7.74&-8.39&-6.81&-6.46&-7.18&\\
$s_a$&-10.66&-11.21&-9.82&-10.58&-11.19&-9.70&-11.47&-11.93&-10.45&\\
$p_a$&3.17&2.06&1.12&3.42&2.43&1.06&2.59&1.45&0.72&\\
$d_a(t_2)$&15.31&16.10&13.08&14.32&15.12&13.20&14.72&15.42&12.76&\\
$d_a(e)$&13.63&14.53&12.29&13.16&14.11&11.62&12.92&13.58&11.66&\\
&&&&&&&&&&\\
$ss\sigma$&-0.73&-0.74&-0.40&-0.72&-0.50&-0.00&-1.01&-0.88&-0.63&-4.08\\
$sp\sigma$&2.57&2.68&2.19&2.12&2.04&2.01&2.05&2.06&1.81&14.49\\
$pp\sigma$&4.95&4.63&3.99&4.40&4.06&3.83&4.16&4.03&3.63&28.01\\
$pp\pi$&-0.88&-0.78&-1.05&-0.44&-0.49&-0.77&-0.48&-0.50&-0.64&-4.51\\
$ps\sigma$&-2.11&-1.28&-1.37&-1.41&-1.70&-0.86&-1.00&-0.90&-1.14&-8.66\\
$ds\sigma$&-0.67&-1.33&-0.32&-0.73&-0.69&-0.35&-1.17&-1.27&-0.74&-5.36\\
$dp\sigma$&0.91&0.84&1.04&0.80&0.67&1.47&0.51&0.93&0.70&5.98\\
$dp\pi$&-0.43&-0.20&-0.10&-0.55&-0.51&-0.17&-0.38&-0.43&-0.37&-2.32\\
$ss\sigma$(2)&-0.10&-0.09&-0.06&-0.04&-0.06&-0.01&-0.00&-0.02&-0.03&-0.30\\
$sp\sigma$(2)&0.32&0.42&0.00&0.01&0.00&0.30&0.21&0.23&0.26&1.30\\
$pp\sigma$(2)&0.61&0.58&0.50&0.48&0.46&0.37&0.52&0.47&0.38&3.23\\
$pp\pi$(2)&-0.06&-0.02&-0.01&-0.03&-0.018&-0.013&-0.03&-0.02&-0.02&-0.16\\
$sd\sigma$&-2.61&-2.87&-2.69&-1.29&-1.44&-1.90&-0.95&-1.47&-1.85&-12.70\\
$pd\sigma$&-3.85&-4.19&-3.98&-2.61&-2.64&-3.46&-2.73&-3.06&-3.06&-22.10\\
$pd\pi$&2.80&2.81&2.55&2.34&2.49&2.24&2.17&2.31&2.07&16.20\\
\hline
\end{tabular}
}
\end{centering}
\end{table}

\begin{thebibliography}{99}
\bibitem{note} The convention we have used for the Slater Koster parameters
indicates the first orbital is on the cation and the second orbital on the
anion. Second nearest-neighbor interactions between anion-anion are marked by
a 2 in the brackets.
\end{thebibliography}
\end{document}